\documentclass[aps,prl,twocolumn,showpacs,floatfix]{revtex4}
\usepackage{xspace}
\usepackage{epsfig}
\usepackage{cases}

\begin{document}
\bibliographystyle{prsty}

\newcommand{\elabel}[1]{\label{eq:#1}}
\newcommand{\eref}[1]{(\ref{eq:#1})}
\newcommand{\Eref}[1]{Eq.~(\ref{eq:#1})}
\newcommand{\flabel}[1]{\label{fig:#1}}
\newcommand{\fref}[1]{Fig.~\ref{fig:#1}}
\newcommand{\Fref}[1]{Fig.~\ref{fig:#1}}
\newcommand{\ave}[1]{{\left\langle #1 \right\rangle}}
\newcommand{\spave}[1]{\overline{#1}}
\newcommand{\GC}{\mathcal{G}}
\newcommand{\CC}{\mathcal{C}}
\newcommand{\vareta}{\xi}
\newcommand{\imag}{i}
\newcommand{\diff}{D}
\newcommand{\half}{\frac{1}{2}}
\newcommand{\comment}[1]{{\bf [#1] }}
\newcommand{\cross}{\mathsf{X}}
\newcommand{\prepsection}[1]{}
\newcommand{\quarter}{\frac{1}{4}}

\title{Drift causes anomalous exponents in growth processes}
\author{Gunnar Pruessner}
\email{gunnar.pruessner@physics.org}
\homepage{http://www.ma.imperial.ac.uk/~pruess/}
\affiliation{
Department of Mathematics,
Imperial College London,
180 Queen's Gate,
London SW7 2BZ,
UK
}
\date{23 October, 2003}

\begin{abstract}The effect of a drift term in the presence of fixed boundaries
is studied for the one-dimensional Edwards-Wilkinson equation, to reveal
a general mechanism that causes a change of exponents for a very broad
class of growth processes. This mechanism represents a relevant
perturbation and therefore is important for the interpretation of
experimental and numerical results. In effect, the mechanism leads to
the roughness exponent assuming the same value as the growth
exponent. In the case of the Edwards-Wilkinson equation this implies
exponents deviating from those expected by dimensional analysis.
\end{abstract}

\maketitle

\prepsection{Introduction}
The Edwards-Wilkinson (EW) equation \cite{EdwardsWilkinson:1982}, as it
is discussed below, is probably the best-studied equation describing
surface growth processes. Due to its linearity it is solvable by
standard methods and has been studied analytically as well as
numerically in great detail
\cite{KrugSpohn:1991,Halpin-HealyZhang:1995,Krug:1997}. The equation is
very well-behaved, so that the outcome of these investigations are
usually quite predictable. There is no reason to suspect that
well-accepted methods, such as dimensional analysis and coarse
graining, produce wrong results, even if applied to the EW equation with
an extra drift term (EWd), which still represents a linear problem.

However, it is shown below that, depending on the boundary conditions,
such a drift term changes the exponents dramatically to anomalous
values, which apparently have been missed in the literature. The drift
in conjunction with the boundary condition poses a relevant perturbation
to the original equation. While it is not possible to capture its effect
by the simple methods mentioned above, it can be understood using
physical arguments. The mechanism turns out to be very powerful and
extends far beyond the EW problem.

\prepsection{The EW equation with drift}
The Edwards-Wilkinson equation \cite{EdwardsWilkinson:1982} describes
the temporal evolution of an interface characterized by its height
$\phi(x,t)$ over a substrate of length $L$, $x \in [0,L]$, at time $t$ under the
influence of a thermal noise $\eta(x,t)$. In one dimension it reads
\begin{equation} \elabel{EW}
 \partial_t \phi(x,t) = \diff \partial_x^2 \phi(x,t) + \eta(x,t) \ ,
\end{equation}
with a diffusion constant or surface tension $\diff$. 
The initial conditions are usually \cite{NattermannTang:1992,Krug:1997}
chosen to be $\phi(x,t=0)\equiv 0$ and periodic boundary conditions
(PBC) are applied. The central property one is interested in is the roughness,
defined as
\begin{equation} \elabel{def_w2}
 w^2(t,L)=\ave{\spave{\phi^2}} - \ave{\spave{\phi}^2} \ ,
\end{equation}
where $\spave{A}$ denotes the spatial average, 
 $\spave{A}=\frac{1}{L} \int_0^L dx A(x)$
and $\ave{\ }$ is the ensemble average, averaging over all realizations of
the noise $\eta$. In order to determine $w^2$, the only property of
$\eta$ which enters is
\begin{equation} \elabel{eta_properties}
 \ave{\eta(x,t)\eta(x',t')}  =  \Gamma^2 \delta(x-x') \delta(t-t') \ ,
\end{equation}
where $\Gamma$ parameterizes the strength of noise. 
To fully specify the noise, the higher order correlations are usually chosen to
be those of Gaussian white noise and the average is set to
$\ave{\eta(x,t)}=0$. Assuming a Family-Vicsek scaling behavior
\cite{FamilyVicsek:1985}, three exponents, $\alpha$, $\beta$ and $z$, are
defined for the asymptotic behavior of $w^2$ 
\begin{equation} \elabel{FV}
 w^2(t,L) = a L^{2\alpha} \GC\left(\frac{t b}{L^z}\right) \ ,
\end{equation}
with appropriate, system dependent parameters $a$ and $b$ which make the
universal scaling function $\GC(x)$ a dimensionless function of a
dimensionless argument. This function is supposed to behave like
$\GC(x) \propto x^{2\beta}$ for small arguments $x$ and to converge to a
non-zero value for large arguments. In the initial growth phase, $w^2$ is
supposedly independent of $L$, so that $\beta=\alpha/z$.

\prepsection{Dimensional Analysis}
Because of the small number of independent parameters, the exponents
characterizing \eref{EW} can be determined immediately, simply by using
dimensional analysis. The roughness based on the ensemble of solutions of
\eref{EW} can be written as 
\begin{equation} \elabel{w2_scaling}
 w^2(t,L; \diff, \Gamma) = \frac{\Gamma^2 L}{\diff} \GC\left(\frac{t \diff}{L^2}\right) \ ,
\end{equation}
with all independent parameters being listed on the left hand
side. \Eref{w2_scaling} is the unique way of writing $w^2$ in the form
prescribed by \eref{FV} with all $t$-dependence being absorbed into a
dimensionless function of a dimensionless argument. By assuming the
existence of the appropriate limits or by simply comparing
\eref{w2_scaling} to \eref{FV}, the exponents are determined to be:
\begin{equation} \elabel{ew_no_drift}
 \alpha=1/2\quad \beta=1/4 \quad z=2\quad \text{(standard EW)}
\end{equation}
Of course, this can easily be confirmed by exact solutions
\cite{NattermannTang:1992,Krug:1997}. 

The exponents \eref{ew_no_drift} remain unchanged, if fixed boundary
conditions (FBC), $\phi(x=0,t)=\phi(x=L,t)=0$, are introduced; the exact
solution changes, but as the new boundary condition does not contain any
new non-zero parameter and therefore cannot introduce any new scale, the
scaling form \Eref{w2_scaling} \emph{must} necessarily remain unchanged;
there are just not enough (independent) parameters for the form
\Eref{w2_scaling} to change.

\prepsection{The drift term}
On the face of it, the EW equation becomes much more complicated, if a
drift or convection term is introduced into \eref{EW},
\begin{equation} \elabel{EWd}
 \partial_t \phi(x,t) = \diff \partial_x^2 \phi(x,t) + v \partial_x \phi(x,t) + \eta(x,t) \ ,
\end{equation}
where $v$ denotes the drifting velocity. Such a term is probably present
in most experimental setups, for example due to the presence of a
gravitational field and a small tilt of the substrate but it is not
obvious whether the resulting drift is strong enough to have a
significant effect, see also the crossover length calculated
below). Here, it is worth mentioning the general r\^ole of a drift term
as discussed in the context of anomalous aging by Luck and Mehta
\cite{LuckMehta:2001}.

In case of PBC, the solution of \Eref{EWd} is related to the solution of
\Eref{EW} by a simple Galilean transformation,
$\phi'(x,t)=\phi(x+vt,t)$, which leaves the noise correlator
\eref{eta_properties} unchanged.
Because of translational
invariance, the correlator $\ave{\phi(x,t)\phi(x',t')}$ (spatially)
depends only on $x-x'+v(t-t')$ and becomes independent of $v$ for equal
times $t'=t$. Because $w^2$ in \eref{def_w2} is a functional of the
\emph{equal-time} correlator, it is also independent of $v$. Thus, in
case of PBC, the exponents remain those listed in \Eref{ew_no_drift},
\begin{equation} \elabel{ewd_pbc}
 \alpha=1/2\quad \beta=1/4 \quad z=2\quad \text{(EW with PBC and drift)}
\end{equation}

Dimensional analysis, however, leads to a scaling form
\begin{equation} \elabel{dimana_drift}
 w^2(t,L; \diff, \Gamma, v) = \frac{\Gamma^2 L}{\diff}\, \GC\left(\frac{t \diff}{L^2}, \frac{vL}{\diff}\right) \ , 
\end{equation}
the behavior of which is unknown; as is shown below, $\GC(x,y)$ might
behave like $y^{-1/2}$ for large $x$. 

\prepsection{Coarse Graining} It is very tempting and in fact leads to
the exponents previously reported in the literature
\cite{BiswasETAL:1998}, to argue that the drift term in \eref{EWd} is
relevant compared to the diffusion term. In fact, this is what an ansatz
$\phi(bx,b^z t)=b^\alpha \phi(x,t)$ suggests. The idea behind this
much-used method \cite{BarabasiStanley:1995,Kardar:1999}, is to
determine the terms dominating the large scale behavior of
\Eref{EWd}. It is often applied to determine relative relevancy of
higher order terms in Langevin equations. In fact, the \emph{exact}
relation
\begin{equation} \elabel{trans}
  \phi(x,t;b L,\diff,\Gamma,v) = b^{\frac{z-1}{2}} \phi(x/b,t/b^z;L,\diff b^{z-2}, \Gamma,v b^{z-1}) 
\end{equation}
seems to indicate that $\phi$ behaves on large scales like $\phi$ on small
scales with a diffusion constant $\diff$ reduced by a factor $b^{-1}$
compared to the reduction of $v$.  Thus, the diffusion constant is
expected to drop out for large system sizes. Repeating the dimensional
analysis without the diffusion term, i.e. setting $\diff=0$, then yields
\begin{equation} \elabel{ewd_coarse}
 \alpha=0\quad \beta=0 \quad z=1\quad \text{(EWd from coarse graining)}\ .
\end{equation}
This is consistent with a scaling law which can be derived from
\eref{trans}. It states that any $\alpha$ derivable from dimensional
analysis must obey \cite{Krug:1997}
\begin{equation} \elabel{dimana_law}
 \alpha=\frac{z-1}{2} \ .
\end{equation}
Below it is shown that this is not the case. There is no reason to
assume that $\phi$ becomes asymptotically self-affine.

As shown above, this result, suggesting a smooth or flat interface
\cite{BiswasETAL:1998}, is wrong at least in the presence of periodic
boundary conditions, because the velocity $v$ simply disappears due to
translational invariance. The key question is then: Are the exponents
\eref{ewd_coarse} recovered if the drift term cannot be gauged away?
Does the coarse graining and dimensional analysis arguments presented
above fail only because of translational invariance, which effectively
removes $v$ as free parameter from the problem? The answer to both of
these questions turns out to be negative. The reason for that is an
extremely efficient mechanism, effectively ``wiping out'' the
stationary roughness.

\prepsection{EW with drift and fixed boundaries}
In order to answer the questions raised above, fixed boundaries are applied
again, so that the drift cannot be gauged away. It is worth stressing
that fixed boundaries correspond to a finite substrate or a finite
region of exposure to the noise, i.e. these boundaries are much more
natural than periodic boundary conditions.

\Eref{EWd} with FBC can be solved using a saddle point approximation. In
the following only a few technical details of this calculation are
presented. After expressing \eref{EWd} in a dimensionless form, the
solution can be written as
\begin{equation} \elabel{formal_solution}
 \varphi(y,\tau;q)=\int_0^1 dy' \int_0^\tau d\tau' \varphi_0(y-y',\tau-\tau') \vareta(y',\tau') \ ,
\end{equation}
where $y=x/L$, $\tau=t/(L^2/\diff)$, $q=v L/\diff$ and 
$\varphi(y,\tau;q) =  1/(\Gamma \sqrt{L/\diff}) \phi(x,t;L,\diff,\Gamma,v)$
as well as
$\vareta(y,\tau)   =  L^{3/2}/(\Gamma \sqrt{\diff})   \eta(x,t)$
are dimensionless quantities. 
The propagator $\varphi_0$ of this problem essentially consists of two
Gaussians ``wrapped around a circle of radius $2$'' from a mirror charge
trick, which are closely related to Jacobi's $\vartheta_3$-function
\cite{FarkasFueloep:2001,MagnusOberhettingerSoni:1966}. This sum is
multiplied by an exponential $e^{-\half (y-y_0)q - \quarter \tau q^2}$,
which accounts for the drift:
\begin{widetext}
\begin{subequations}
\begin{eqnarray} 
 \varphi_0(y,\tau;y_0,q) & = &
\frac{1}{\sqrt{4 \pi \tau}}
\sum_{n=-\infty}^\infty \left(
e^{-\frac{(y - y_0 + 2 n)^2}{4 \tau}}-e^{-\frac{(y + y_0 + 2 n)^2}{4 \tau}}
\right)
 e^{-\half (y-y_0)q - \quarter \tau q^2}  \\ 
&= & 
2 \sum_{n=1}^{\infty} \sin(n \pi y) \sin(n \pi y_0) e^{-(n \pi)^2 \tau}  
 e^{-\half (y-y_0)q - \quarter \tau q^2} 
\end{eqnarray}
\end{subequations}
where by convention $\tau_2\ge\tau_1$ and $y_0$ is the initial
position. The saddle-point approximation is required when the spatial
averages \eref{def_w2} are taken. One finds
\begin{subnumcases}{\elabel{w2_fbc} w^2(t,L)  =  \Gamma^2 \times}
\sqrt{\frac{t}{2\pi \diff}} - \frac{t}{L} - \frac{|v| t ^{3/2}}{3 L \sqrt{2 \pi \diff}} 
+ \frac{t^2 |v|}{2L^2} & for $t\ll \frac{L}{|v|} \qquad$
 \elabel{w2_fbc_initial}\\
\frac{2}{3 \sqrt{2\pi}}
 \sqrt{\frac{L}{|v|\diff}} - \frac{1}{2 |v|} & for $t\gg \frac{L}{|v|} \qquad$ \ ,
\elabel{w2_fbc_terminal}
\end{subnumcases}
\end{widetext}
which becomes exact for divergent $vL/D$, i.e. especially in the
thermodynamic limit. Inspecting the leading terms immediately gives
\begin{equation} \elabel{ewd_proper}
 \alpha=1/4\quad \beta=1/4 \quad z=1\quad \text{(EW with FBC and drift)}
\end{equation}
This result is surprising, because these exponents are \emph{anomalous},
as they do not correspond to what seems to be suggested by dimensional
analysis: Not only do they contradict \eref{ew_no_drift}, \eref{ewd_pbc}
and \eref{ewd_coarse}, they do not even obey
\eref{dimana_law}. In fact, neither the drift term, nor the diffusion
term become irrelevant\footnote{Field-theoretically, the diffusion term
might be regarded as dangerously irrelevant.}; they are both crucial for
\eref{ewd_proper}. Moreover, the set of exponents switches from
\eref{ewd_pbc} to \eref{ewd_proper} if the boundary condition are
changed from periodic to fixed in the presence of a drift term.
The Edwards-Wilkinson equation with drift
term and fixed boundary conditions is a \emph{linear} problem displaying
\emph{anomalous} exponents. 


\prepsection{Physical explanation}
The fact that the above result cannot be reconciled with dimensional analysis
might suggest that the mechanism leading to these exponents is very
subtle. However, it turns out that it can be understood quite easily.

\begin{figure}[t]
\begin{center}
\includegraphics[width=0.9\linewidth]{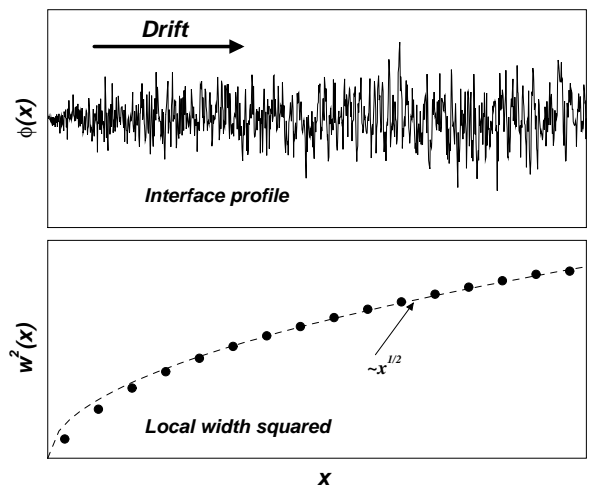}
\caption{\flabel{reinit_fig}A qualitative picture of an interface
snapshot with its ``local roughness'', obtained in a numerical simulation; scales are irrelevant. Upper
panel: An example of an interface profile with fixed boundaries and
drift term. Lower panel: The ensemble averaged \emph{local} width
squared (numerical data, circles) is proportional to the square root of
the position where measured, $x^{1/2}$ (fitted, dashed line).  }
\end{center}
\end{figure}

The drift term makes the entire interface configuration move from one
boundary to the other. Without noise, a peak starting somewhere in the
bulk gets slowly moved by the drift to one of the boundaries, while
diffusively broadening. It eventually disappears at the boundary. The
time it spends between the boundaries depends on the starting position
and the direction of the movement. The maximum time is $L/v$, which is
also the maximum time, any noise-generated structure has to
develop. However, as known from the model without drift, it takes time
$L^2/\diff$ in order to fully develop the roughness. Thus, for
$L^2/\diff \gg L/v$, i.e. $L\gg \diff/v=L_\cross$, the interface will
remain in its initial growth phase; the characteristic
length scale $L_\cross$ represents an effective cutoff for
correlations. At the same time it enables the system to display
anomalous exponents. Depending on the direction of the drift, the interface ``comes out''
of the left boundary initialized to $\phi=0$ and moves to the right
boundary, as shown in \fref{reinit_fig}. The average ``age'' is
proportional to $t_\cross=L/v$, so that according to \eref{FV}
$w^2\propto t_\cross^{1/2}\propto L^{1/2}$, therefore
$\alpha=1/4$. Regarding $\beta$, the interface cannot ``see'' the drift
initially, so that $\beta=1/4$ just like for the case without drift,
\eref{ew_no_drift}. Indeed, even the amplitude of the leading term in
\eref{w2_fbc_initial} corresponds to the amplitude obtained for the
problem with periodic boundaries. The identity $\alpha=\beta$ already
indicates $z=1$, which can also be derived from the fact that saturation
should be reached as soon as the interface has swiped through the system
once, i.e. after $t_\cross=L/v$.

To test the validity of the results, \Eref{w2_fbc} has been compared to
numerical simulations, based on a straight-forward integration of
\eref{EWd} using Euler's method. For system sizes not too small ($L\ge 128$),
even the higher order corrections were reproduced. The mechanism is
illustrated in \Fref{reinit_fig}: The upper panel shows a snapshot of an
interface configuration. The ``local age'' of the interface can be read
off the local roughness (in an appropriate \emph{ad hoc} definition) as
shown in the panel below, because spending more time between the
boundaries increases the local roughness according to $w^2\propto
t^{1/2}= (x/v)^{1/2}$, with $x$ being the position where the roughness
is measured.

\prepsection{Discussion and conclusion}
The physical explanation presented above goes beyond the EW equation;
provided that the crossover time of the original model without drift
scales faster in $L$ than $t_\cross$, i.e. $z>1$, the argument should
apply, so that at sufficiently large system sizes $\alpha$ obtains the
value of $\beta$ and therefore $z=1$. It is a very efficient mechanism,
which works under very general circumstances even in the most simple,
linear case. It therefore speaks a clear warning as to the
interpretation of numerical and experimental studies: the true value of
$\alpha$ might have been ``washed away'' by a very small drift.

The KPZ
\cite{KardarParisiZhang:1986,KrugSpohn:1991,Halpin-HealyZhang:1995,Krug:1997}
equation ($z=3/2$) is particularly interesting: Its non-linearity is
only important during the initial growth phase ($\beta=1/3$) and becomes
insignificant in the stationary regime ($\alpha=1/2$). However, with an
additional drift term the equation should remain in the initial growth
phase, with the non-linearity present in the stationary state. Indeed,
using a Cole-Hopf transformation \cite{Halpin-HealyZhang:1995}, the
problem can be reduced to an equation similar to \Eref{EWd}. However,
preliminary numerical tests did not fully confirm this correspondence.

It is not yet completely clear how to generalize this arguments to
higher dimensions. For two dimensions it is tempting to speculate
whether exponents observed in experimental molecular beam epitaxy are
related to such a drift term, for example when $\alpha \approx \beta$
\cite{YouETAL:1993} or when $\alpha$ is close to
$\beta_{\text{KPZ}}\approx 0.24$ \cite{Kardar:1999,EklundETAL:1991}.
Interestingly, only one boundary needs to be fixed in order to observe
the phenomenon, namely the boundary the velocity points away from, in
\Fref{reinit_fig} the left boundary.

The mechanism stresses once more the relevance of boundary conditions as
prominently pointed out by Landau and Binder
\cite{LandauBinder:1988}. However, it is worth emphasizing that in the
present case, the change of boundary conditions leads to a change of the
\emph{bulk} critical exponents.

Even though the physical explanation provides a very clear
picture of the mechanism, it is not obvious how a drift term
exactly affects other models, with, for example, quenched or
conserved noise, with an additional term $-m\phi$ or with
additional scaling laws.


In conclusion we have presented a remarkably simple mechanism which
reduces the roughness exponent to the value of the growth exponent for
any small amount of drift in the Langevin equation in the presence of
fixed boundary conditions, provided that in the original model, the
dynamical exponent $z$ is larger than unity. On sufficiently large
scale, this mechanism should be visible in many experimental and
numerical systems. Most unexpectedly, it can even be found in the
Edwards-Wilkinson equation, which consequently shows anomalous exponents,
depending on the boundary conditions imposed.

\acknowledgments The author would like to thank {\'E}. Brunet, E. Frey,
H. J. Jensen, J. M. L{\'o}pez, J. M. Luck, A. Parry, B. Schmittmann, U. C. T{\"a}uber and
R. K. P. Zia for very helpful discussions and for their
hospitality. Also, he thanks H. Bruhn, C. Porter, I. Ruf, the Thoeren
family, and especially the Max-Planck-Institut f{\"u}r Physik komplexer
Systems in Dresden for hospitality during NESPHY03. The author
gratefully acknowledges the support of EPSRC.

\emph{Note added in proof:} The author wants to thank J. Krug for
pointing out that the effect of a drift term is discussed and referenced
in Sec 4.6.4 of \cite{Krug:1997} in relation to KPZ, see especially the
work by Derrida et al. \cite{DerridaETAL:1991}, also with respect to the
EW equation \cite{DerridaETAL:1991b}.

\bibliography{articles,books}
\end{document}